\documentclass[11pt]{article}

\usepackage{amsmath,amssymb}
\usepackage[active]{srcltx}
\usepackage{hyperref}
\def\tabaddress#1{{\small\it\begin{tabular}[t]{c}#1
\\[1.2ex]\end{tabular}}}
\def\UPCMAT{Dep. Mathematics.
   Campus Nord U.P.C., Ed. C-3\\
   C/ Jordi Girona 1.
   E-08034 Barcelona, SPAIN}

\font\fr=eufm10 scaled \magstep 1 

\newtheorem{definition}{Definition}

\def\beq{\begin{equation}}
\def\eeq{\end{equation}}
\def\bea{\begin{eqnarray}}
\def\eea{\end{eqnarray}}
\def\beann{\begin{eqnarray*}}
\def\eeann{\end{eqnarray*}}
\def\ben{\begin{enumerate}}
\def\een{\end{enumerate}}
\def\bit{\begin{itemize}}
\def\eit{\end{itemize}}
\def\derpar#1#2{\frac{\partial{#1}}{\partial{#2}}}

\def\qed{\ifvmode\removelastskip\fi
{\unskip\nobreak\hfil\penalty50\hbox{}\nobreak\hfil \hbox{\vrule
height1.2ex width1.2ex}\parfillskip=0pt \finalhyphendemerits=0
\par\smallskip}}
\def\vf{\mbox{\fr X}}

\def\d{{\rm d}}

\def\inn{\mathop{i}\nolimits}

\def\Cinfty{{\rm C}^\infty}

\title{MULTISYMPLECTIC FORMULATION OF LAGRANGIAN MODELS IN GRAVITATION}
\author{\sc Jordi Gaset,
Narciso Rom\'an-Roy
 \\
   \tabaddress{\UPCMAT}\\
{\small ({\bf e}-{\it mails}: jordi.gaset@upc.edu,
narciso.roman@upc.edu)}}
   \date{September 24, 2019}

\begin{document}

\maketitle

\begin{abstract}
We apply the multisymplectic formulation of classical field theories \cite{pere,art:Roman09,book:Saunders89} to describe the Einstein-Hilbert and the Einstein-Palatini (or metric-affine) Lagrangian models of General Relativity.
\end{abstract}

\begin{small}
\begin{center}


\bigskip
\vbox{AMS s.\,c.\,(2010): \null 
{\it Primary}: 53D42, 55R10, 70S05, 83C05; {\it Secondary}: 49S05, 53C15, 53Z05.}\null

\end{center}
\end{small}

\section{Introduction}

The geometrization of the theory of gravity ({\sl General Relativity} (GR)) and, 
in particular, the {\sl multisymplectic framework}, 
allows us to
understand several inherent characteristics of it.
It is studied by different authors, such as
\cite{art:Capriotti,art:Capriotti2,first,Ka-01,Ka-13,
Krupka,KrupkaStepanova,rosado2,vey1}.

We present the main Lagrangian models for GR 
using the  {\sl multisymplectic framework}:
first the {\sl Einstein-Hilbert model} which
is described by a $2nd$-order singular Lagrangian
(and so GR is formulated
as a higher-order premultisymplectic field theory with constraints),
and second the {\sl Einstein-Palatini (metric-affine) model}
described by a $1st$-order singular Lagrangian
(and so GR is formulated
as a 1st-order premultisymplectic field theory with constraints).

\section{Geometric structures: Jet bundles and multivector fields}

First we introduce some fundamental geometrical tools which are used in the exposition.

Let $\pi\colon E\longrightarrow M$ be a fiber bundle
(with adapted coordinates $(x^\mu,y^i)$).
A  {\sl section} of $\pi$ is a map $\phi\colon U\subset M\to E$ such that $\pi\circ\phi=Id_M$.
The set of sections is denoted $\Gamma(\pi)$.
Two sections $\phi_1,\phi_2\in\Gamma(\pi)$ are {\sl $k$-equivalent}
at $x\in M$ if $\phi_1(x)=\phi_2(x)$ and their partial derivatives 
until order $k$ at $x$ are equal.
This is an {\sl equivalence relation} in  $\Gamma_x(\pi)$
and each equivalence class
is a {\sl jet field} at $x$; denoted $j^k\phi_x$.
The {\sl $k$th-order jet bundle} of $\pi$ is the set
 $J^k\pi:=\{ j^k_x\phi\, \vert \, x\in M,\, \phi\in\Gamma_x(\pi)\}$.
 Natural projections are:
$$ 
\pi^k_r \colon J^k\pi  \longrightarrow J^r\pi  \  \ (r<k)\quad , \quad
\pi^k \colon J^k\pi \longrightarrow E  \quad , \quad
\bar{\pi}^k \colon J^k\pi \longrightarrow M \ .
$$
\begin{definition}
The {\bf $k$th-prolongation} of a section $\phi\in\Gamma(\pi)$ to $J^k\pi$
is the section $j^k\phi\in\Gamma(\bar\pi^k)$ defined as
$j^k\phi(x):=j^k_x\phi\ ; \ x\in M$.
A section $\psi \in \Gamma(\bar{\pi}^{k})$ in $J^k\pi$ is 
{\bf holonomic} if
$\psi=j^k\phi$; that is, $\psi$ is the $k$th prolongation of a section
$\phi = \pi^{k} \circ \psi \in \Gamma(\pi)$.
\end{definition}
If $\phi=(x,y^i(x))$, then 
$\displaystyle\psi=j^k\phi=\Big(x,y^i(x),\derpar{y^i}{x^\mu}(x),\frac{\partial^2y^i}{\partial x^\mu\partial x^\nu}(x),\ldots\Big)$.
\begin{definition}
An {\bf $m$-multivector field}  in $J^k\pi$ is a skew-symmetric contravariant 
tensor of order $m$ in $J^k\pi$. The set of $m$-multivector fields 
in $J^k\pi$ is denoted $\vf^m (J^k\pi)$.
A multivector field $\mathbf{X}\in\vf^m(J^k\pi)$ is said to be 
{\bf locally decomposable} if,
for every $p\in J^k\pi$, there is an open neighbourhood  $U_p\subset J^k\pi$
and $X_1,\ldots ,X_m\in\vf (U_p)$ such that $\mathbf{X}\vert_{U_p}=X_1\wedge\ldots\wedge X_m$.
Locally decomposable $m$-multivector fields $\mathbf{X}\in\vf^m(J^k\pi)$ 
are locally associated with $m$-dimensional
distributions $D\subset{\rm T}J^k\pi$. Then,
$\mathbf{X}$ is {\bf integrable} if its associated distribution is integrable. 
In particular,
$\mathbf{X}$ is {\bf holonomic} if
it is integrable and 
its integral sections are holonomic sections of $\bar\pi^k$.
\end{definition}

If $\Omega\in\Omega^r(J^k\pi)$ is a differential $r$-form in $J^k\pi$ and $\mathbf{X}\in\mathfrak{X}^m(J^k\pi)$ is locally decomposable,
the {\sl contraction} between ${\bf X}$ and 
$\Omega$ is
$ \inn({\bf X})\Omega\mid_{U}:= \inn(X_1)\ldots\inn(X_m)\Omega$.

 \section{Einstein-Hilbert model (without sources)}

The {\sl configuration bundle} for the Einstein-Hilbert model
is $\pi\colon E\rightarrow M$, where
$M$ is an oriented, connected 4-dimensional manifold representing space-time, with
volume form $\omega \in\Omega^4(M)$, and
$E$ is the manifold of {\sl Lorentzian metrics} on $M$. Thus $\dim E=14$.
Adapted fiber coordinates in $E$ are $(x^\mu,g_{\alpha\beta})$,
(with $0\leq\alpha\leq\beta\leq 3$),
where $g_{\alpha\beta}$ are the components of the metric, and
such that 
$\omega=\d x^0\wedge\d x^1\wedge\d x^2\wedge\d x^3\equiv\d^4x$.
(We also use the notation
$\displaystyle \d^3x_\mu\equiv\inn\Big(\derpar{}{x^\mu}\Big)\d^4x$).

The Lagrangian formalism is developed in $J^3\pi$,
with the induced coordinates denoted as
$(x^\mu,\,g_{\alpha\beta},\,g_{\alpha\beta,\mu},\,g_{\alpha\beta,\mu\nu},
\,g_{\alpha\beta,\mu\nu\lambda})$, ($0\leq\alpha\leq\beta\leq 3$;
$0\leq\mu\leq\nu\leq\lambda\leq 3$).
The bundle $J^3\pi$ has some canonical structures; in particular, the {\sl total derivatives} are
$$
\displaystyle
D_\tau=\derpar{}{x^\tau}
+ g_{\alpha\beta,\tau}\derpar{}{g_{\alpha\beta}}+ g_{\alpha\beta,\mu\tau}\derpar{}{g_{\alpha\beta,\mu}}+ g_{\alpha\beta,\mu\nu\tau}\derpar{}{g_{\alpha\beta,\mu\nu}}
+ g_{\alpha\beta,\mu\nu\lambda\tau}\derpar{}{g_{\alpha\beta,\mu\nu\lambda}} \ .
$$

The {\sl Hilbert-Einstein Lagrangian function} (without energy-matter) is
$$
L_{EH}=\sqrt{|det(g_{\alpha\beta})|}R\equiv\varrho R=\varrho g^{\alpha\beta}R_{\alpha\beta}\in\Cinfty(J^2\pi) \ ;
$$
where $R_{\alpha\beta}$ are the {\sl Ricci tensor} components,
$\displaystyle\Gamma^{\rho}_{\mu\nu}$
are the {\sl Christoffel symbols} of the {\sl Levi-Civita connection} of $g$,
and $R$ is the {\sl scalar curvature} 
(which contains $2nd$-order derivatives of $g_{\mu\nu}$).
The {\sl Hilbert-Einstein Lagrangian density} is
$\mathcal{L}=L\,\d^4x\in\Omega^4(J^3\pi)$,  where
$L=(\pi^3_2)^*L_{EH}\in C^\infty(J^3\pi)$.
We denote
\beann
L^{\alpha\beta,\mu\nu}&=&\displaystyle
\frac{1}{n(\mu\nu)}\frac{\partial L}{\partial g_{\alpha\beta,\mu\nu}}=\frac{n(\alpha\beta)}{2}\varrho(g^{\alpha\mu}g^{\beta\nu}+g^{\alpha\nu}g^{\beta\mu}-2g^{\alpha\beta}g^{\mu\nu})\ ,
 \\
L^{\alpha\beta,\mu}&=&\displaystyle  
\frac{\partial L}{\partial g_{\alpha\beta,\mu}} - \sum_{\nu=0}^{3}\frac{1}{n(\mu\nu)}D_\nu\Big( \frac{\partial L}{\partial g_{\alpha\beta,\mu\nu}}\Big)=\frac{\partial L}{\partial g_{\alpha\beta,\mu}} - \sum_{\nu=0}^{3}D_\nu L^{\alpha\beta,\mu\nu}\ .
\\
H&=&
\displaystyle \sum_{\alpha\leq \beta ; \mu\leq \nu}L^{\alpha\beta,\mu\nu}g_{\alpha\beta,\mu\nu}+
\sum_{\substack{\alpha\leq \beta}}L^{\alpha\beta,\mu}g_{\alpha\beta,\mu}-L=
\varrho\, g_{\alpha\beta,\mu}g_{kl,\nu}H^{\alpha\beta k l \mu\nu} \ , 
\\
H^{\alpha\beta k l\mu\nu}&=&
\frac{1}{4}g^{\alpha\beta}g^{kl}g^{\mu\nu}-\frac{1}{4}g^{\alpha k}g^{\beta l}g^{\mu\nu}+\frac12g^{\alpha k}g^{l\mu}g^{\beta\nu}-\frac12g^{\alpha\beta}g^{l\nu}g^{k\mu} \ ,
\eeann
(where $n(\mu\nu)=1$ if $\mu=\nu$, and $n(\mu\nu)=2$ if $\mu\neq\nu$).
Then, the {\sl Poincar\'e-Cartan $5$-form} associated with ${\mathcal L}$ is
$$
\Omega_{\mathcal{L}}=
\d H\wedge\d^4x
-\sum_{\alpha\leq\beta}\d L^{\alpha\beta,\mu}\d g_{\alpha\beta}\wedge \d^{m-1}x_\mu-\sum_{\alpha\leq\beta}\d L^{\alpha\beta,\mu\nu}\d g_{\alpha\beta,\mu}\wedge \d^{m-1}x_{\nu}\in\Omega^5(J^3\pi) \ ,
$$
and it is a premultisymplectic form because $L$ is a singular Lagrangian.

The problem stated by the {\sl Hamilton variational principle} for the system 
$(J^3\pi,\Omega_\mathcal{L})$
consists in finding holonomic sections 
$\psi_\mathcal{L}=j^3\phi\in \Gamma(\bar\pi^3)$ 
satisfying any of the following equivalent conditions:
\begin{description}
\item[{\rm (a)}] $\psi_\mathcal{L}$ is a solution to the equation
$\psi_\mathcal{L}^*\inn(X)\Omega_\mathcal{L} = 0$,
for every $X \in \mathfrak{X}(J^3\pi)$.
\item[{\rm (b)}] $\psi_\mathcal{L}$ is an integral section of a holonomic multivector field ${\bf X}_\mathcal{L}\in\mathfrak{X}^4(J^3\pi)$
satisfying the equation
$\inn({\bf X}_\mathcal{L})\Omega_\mathcal{L} = 0$.
\end{description}
As $\Omega_{\mathcal{L}}$ is a premultisymplectic form, these field equations have no solution everywhere in $J^3\pi$.
Applying the {\sl premultisymplectic constraint algorithm} we obtain the following constraints (see  \cite{GR1}):
\bea
L^{\alpha\beta}:=
-\varrho\,n(\alpha\beta) (R^{\alpha\beta}-\frac{1}{2}g^{\alpha\beta}R)=0 \ .
\label{lagcons1}
\\
D_\tau L^{\alpha\beta}=
D_\tau(-\varrho\,n(\alpha\beta)(R^{\alpha\beta}-\frac{1}{2}g^{\alpha\beta}R))=0 \ .
\label{lagcons2}
\eea
They define the Lagrangian final constraint submanifold
$S_f\hookrightarrow J^3\pi$  where solutions exist and, in particular,
\beann
{\bf{X}}_\mathcal{L}=&
\displaystyle\bigwedge_{\tau=0}^3 \sum_{\alpha\leq\beta}\sum_{\mu\leq\nu\leq\lambda}\Big(\derpar{}{x^\tau}+ g_{\alpha\beta,\tau}\frac{\partial}{\partial g_{\alpha\beta}}+ g_{\alpha\beta,\mu\tau}\frac{\partial}{\partial g_{\alpha\beta,\mu}}+
\\
&g_{\alpha\beta,\mu\nu\tau}\derpar{}{g_{\alpha\beta,\mu\nu}}+
D_\tau D_\lambda (g_{\lambda\sigma}(\Gamma_{\nu \alpha }^\lambda\Gamma_{\mu \beta}^\sigma+\Gamma_{\nu \beta}^\lambda\Gamma_{\mu \alpha }^\sigma))\derpar{}{g_{\alpha\beta,\mu\nu\lambda}}\Big) 
\eeann
is a holonomic multivector field solution to 
the equation in (b), tangent to $S_f$.
Their integral sections are the solutions
$\psi_\mathcal{L}(x)=(x^\mu,\,g_{\alpha\beta}(x),\,g_{\alpha\beta,\mu}(x),\,g_{\alpha\beta,\mu\nu}(x),
\,g_{\alpha\beta,\mu\nu\lambda}(x))$ 
to the equation in (a), which gives 
\bea
g_{\alpha\beta,\mu}-\frac{\partial g_{\alpha\beta}}{\partial x^\mu}&=&0 \ ,
\label{eqsecunif4} \\
g_{\alpha\beta,\mu\nu}-\frac{1}{n(\mu\nu)}\Big(\frac{\partial g_{\alpha\beta,\mu}}{\partial x^{\nu}}+\frac{\partial g_{\alpha\beta,\nu}}{\partial x^{\mu}}\Big)&=&0 \ ,
\label{eqsecunif5}\\
\varrho\,n(\alpha\beta)(R^{\alpha\beta}-
\frac{1}{2}g^{\alpha\beta}R)&=&0 \ .
\label{eqsecunif1}
\eea

In this set of equations, \eqref{eqsecunif4} and \eqref{eqsecunif5}
are (part of the) holonomy conditions;
meanwhile (\ref{eqsecunif1})
are the physical relevant equations, which
are the constraints \eqref{lagcons1} evaluated on the image of sections,
$L^{\alpha\beta}|_{\psi_\mathcal{L}}=0$,
and constitute the Euler-Lagrange equations of the theory;
that is, the {\sl Einstein equations}.

As a consequence of the singularity of $\mathcal{L}$,
the form $\Theta_\mathcal{L}$ is $\pi^3_1$-projectable onto a form in $J^1\pi$
(but it is not the Poincar\'e-Cartan form of any 1st-order Lagrangian). 
Then,
Einstein equations are $2nd$-order PDE's, instead of $4th$-order 
as it correspond to a $2nd$-order Lagrangian.
So they are defined as a submanifold of $J^3\pi$
(and appear as constraints).

The constraints  \eqref{lagcons2} are of geometrical nature and arise because we are using a manifold 
prepared for a theory of a $2nd$-order Lagrangian that, really,
is physically equivalent to a $1st$-order Lagrangian.
These constraints hold automatically when they are evaluated on the image of 
the sections $\psi_\mathcal{L}$ which are solutions to the Einstein equations.
Furthermore, the Einstein-Hilbert model is a gauge theory (because $L_{EH}$ is singular). 
Then, the constraints \eqref{lagcons1} and \eqref{lagcons2} fix partially the gauge.
To remove the remaining gauge degrees of freedom leads to 
a submanifold of $S_f$ diffeomorphic to $J^1\pi$.

\section{Einstein-Palatini (metric-affine) model (without sources)}

The configuration bundle of the Einstein-Palatini (metric-affine) model is $\Pi\colon{\mathcal E}\rightarrow M$, where
${\mathcal E}=E\times_MC(LM)$,
where $E$ is the manifold of Lorentzian metrics on $M$ and
$C(LM)$ is the manifold of {\sl linear connections} in ${\rm T}M$.
Adapted fiber coordinates in $E$ are 
$(x^\mu,g_{\alpha\beta},\Gamma_{\mu\nu}^\lambda)$,
($0\leq\alpha\leq\beta\leq 3$), and
the induced coordinates in $J^1\Pi$ are 
$(x^\mu,g_{\alpha\beta},\, \Gamma_{\mu\nu}^\lambda,\,g_{\alpha\beta,\mu},\,\Gamma_{\mu\nu,\rho}^\lambda)$.
Thus $\dim E=78$ and $\dim J^1\Pi=374$.

The {\sl Einstein-Palatini Lagrangian}
(without energy-matter) is a singular $1st$-order Lagrangian
depending on the components of the metric $g$ 
and of a connection $\Gamma$,
$$
L_{EP}=\varrho\,g^{\alpha\beta}R_{\alpha\beta}=
\varrho\,g^{\alpha\beta}(\Gamma^{\gamma}_{\beta\alpha,\gamma}-\Gamma^{\gamma}_{\gamma\alpha,\beta}+
\Gamma^{\gamma}_{\beta\alpha}\Gamma^{\sigma}_{\sigma\gamma}-
\Gamma^{\gamma}_{\beta\sigma}\Gamma^{\sigma}_{\gamma\alpha})\in\Cinfty(J^1\Pi) \ .
$$
The Lagrangian density is 
${\mathfrak L}=L_{EP}\,\d^4x\in \Omega^4(J^1\Pi)$,
and its {\sl Poincar\'e-Cartan $5$-form} is
$$
\Omega_{\mathfrak L}=
\d\Big(\frac{\partial L_{\rm EP}}{\partial \Gamma^{\alpha}_{\beta\gamma,\mu}}\Gamma^{\alpha}_{\beta\gamma,\mu}-L_{\rm EP}\Big)\wedge\d^4x -
\d\frac{\partial L_{\rm EP}}{\partial \Gamma^{\alpha}_{\beta\gamma,\mu}}\wedge\d \Gamma^{\alpha}_{\beta\gamma}\wedge \d^3x_{\mu}\in \Omega^5(J^1\Pi) \ ,
$$
which is a premultisymplectic form since $L_{EP}$ is also a singular Lagrangian.

The {\sl Lagrangian problem} for the system 
$(J^1\Pi,\Omega_\mathfrak{L})$ consists in
finding holonomic sections $\psi_\mathfrak{L}=j^1\phi\in \Gamma(\bar\pi^1)$ ($\phi\in\Gamma(\Pi)$)
satisfying any of the following equivalent conditions:
\begin{description}
\item[{\rm (a)}] 
$\psi_\mathfrak{L}$ is a solution to the equation
$\psi_\mathfrak{L}^*\inn(X)\Omega_\mathfrak{L} = 0 $,
for every $X \in \mathfrak{X}(J^1\Pi)$.
\item[{\rm (b)}]
$\psi_\mathfrak{L}$ is an integral section of a holonomic multivector field ${\bf X}_\mathcal{L}\in\mathfrak{X}^4(J^1\Pi)$
satisfying the equation
$\inn({\bf X}_\mathfrak{L})\Omega_\mathfrak{L} = 0$.
\end{description}

Now, the premultisymplectic constraint algorithm leads to the constraints (see  \cite{GR2}):
\bea
0&=&\frac{\partial H}{\partial g_{\mu\nu}}-
\frac{\partial L_\alpha^{\beta\gamma,\sigma}}{\partial g_{\mu\nu}}\Gamma^\alpha_{\beta\gamma,\sigma} \ ,
\label{cero}
\\
0&=&g_{\rho\sigma,\mu}-g_{\sigma\lambda}\Gamma^\lambda_{\mu\rho}-
g_{\rho\lambda}\Gamma^\lambda_{\mu\sigma}-\frac{2}{3}g_{\rho\sigma}T^\lambda_{\lambda\mu} \ ,
\label{uno}
\\
0&=&T^\alpha_{\beta\gamma}-
\frac13\delta^\alpha_\beta T^\mu_{\mu\gamma}+
\frac13\delta^\alpha_\gamma T^\mu_{\mu\beta} \ ,
\label{dos}
\\
0&=&T^\alpha_{\beta\gamma,\nu}-
\frac13\delta^\alpha_\beta T^\mu_{\mu\gamma,\nu}+
\frac13\delta^\alpha_\gamma T^\mu_{\mu\beta,\nu} \ ,
\label{tres}
\\
0&=&g_{\rho\gamma}\Gamma^\gamma_{[\nu\lambda}\Gamma^\lambda_{\mu]\sigma}+g_{\sigma\gamma}\Gamma^\gamma_{[\nu\lambda}\Gamma^\lambda_{\mu]\rho}+g_{\rho\lambda}\Gamma^\lambda_{[\mu\sigma,\nu]}+g_{\sigma\lambda}\Gamma^\lambda_{[\mu\rho,\nu]}+\frac23g_{\rho\sigma}T^\lambda_{\lambda[\mu,\nu]}\ . 
\label{cuatro}
\eea
where 
$T^\alpha_{\beta\gamma}\equiv\Gamma^\alpha_{\beta\gamma}-\Gamma^\alpha_{\gamma\beta}$.
They define the  submanifold ${\mathcal S}_f\hookrightarrow J^1\Pi$,
where there are holonomic multivector fields solution to
the equations in (b), tangent  to ${\mathcal S}_f$.

A consequence of the singularity of $\mathfrak{L}$ is that
$\Omega_\mathfrak{L}$ is $\Pi^1$-projectable onto a form in ${\mathcal E}$
and then,  the Euler-Lagrange equations
(Einstein's eqs.) are $1st$-order PDE's, instead of $2nd$-order.
So they are defined as a submanifold of $J^1\pi$,
and appear as constraints \eqref{cero}.
On the other hand, the equalities  \eqref{uno}
are related to the metricity condition for the Levi-Civita connection and they are called {\sl pre-metricity constraints}.
Furthermore, there are the {\sl torsion constraints} 
that impose conditions on the torsion of the connection
\eqref{dos}
and on their derivatives  \eqref{tres}.
Finally, the additional {\sl  integrability constraints} \eqref{cuatro}
appear as a consequence of demanding the integrability of the multivector fields 
which are solutions to the equations in (b).

The Einstein-Palatini model is a gauge theory (as ${\mathfrak L}$ is singular)
with higher gauge freedom than in the Einstein-Hilbert  model.
The above constraints fix partially the gauge.
To remove the remaining gauge degrees of freedom leads to 
a submanifold of ${\mathcal S}_f$ diffeomorphic to $J^1\pi$ in the Einstein-Hilbert model.
The conditions of the connection to be  
{\sl torsionless} and {\sl metric}
(which allows us to recover the Einstein-Hilbert model from the Einstein-Palatini model)
are a consequence of the constraints and a partial fixing of this gauge freedom \cite{pons}.

\begin{small}

\section*{Acknowledgments}

We acknowledge the financial support from the 
Spanish Ministerio de Econom\'{\i}a y Competitividad
project MTM2014--54855--P, 
the Ministerio de Ciencia, Innovaci\'on y Universidades project
PGC2018-098265-B-C33,
and the Secretary of University and Research of the Ministry of Business and Knowledge of
the Catalan Government project
2017--SGR--932.

\end{small}

\end{document}